\documentclass[10pt,twocolumn,letterpaper]{article}

\usepackage{cvpr}              

\def\eg{\emph{e.g.}}

\def\ie{\emph{i.e.}}

\definecolor{cvprblue}{rgb}{0.21,0.49,0.74}
\usepackage[pagebackref,breaklinks,colorlinks,allcolors=cvprblue]{hyperref}
\def\paperID{********}
\def\confName{CVPR}
\def\confYear{2025}

\title{EmoDubber: Towards High Quality and Emotion Controllable Movie Dubbing}

\begin{document}

\author{Gaoxiang Cong$^{1,2}$\footnotemark[1]~~~Jiadong Pan$^{1,2}$\footnotemark[1]~~~Liang Li$^{1}$\footnotemark[2]~~~Yuankai Qi$^{3}$~~~Yuxin Peng$^{4}$\\~~~Anton van den Hengel$^{5}$
~~~Jian Yang$^{3}$~~~Qingming Huang$^{1,2}$\\
	$^1$Institute of Computing Technology, Chinese Academy of Sciences~~~$^3$Macquarie University\\
    $^2$University of Chinese Academy of Sciences~~~$^4$Peking University~~~$^5$University of Adelaide\\
    }

\maketitle

\renewcommand{\thefootnote}{\fnsymbol{footnote}}
\footnotetext[1]{Equal contribution.}
\footnotetext[2]{Corresponding author.}
\renewcommand{\thefootnote}

\begin{abstract}

Given a piece of text, a video clip, and a reference audio, the movie dubbing 
task aims to generate speech that aligns with the video while cloning the desired voice. 
The existing methods have two primary deficiencies: (1) They struggle to simultaneously hold audio-visual sync and achieve clear pronunciation; (2) They lack the capacity to express user-defined emotions. 
To address these problems, we propose EmoDubber, an emotion-controllable dubbing architecture that allows users to specify emotion type and emotional intensity while satisfying high-quality lip sync and pronunciation. 
Specifically, we first design Lip-related Prosody Aligning (LPA), which focuses on learning the inherent consistency between lip motion and prosody variation by duration level contrastive learning to incorporate reasonable alignment. 
Then, we design Pronunciation Enhancing (PE) strategy to fuse the video-level phoneme sequences by efficient conformer to improve speech intelligibility. 
Next, the speaker identity adapting module decodes acoustics prior and inject the speaker style embedding. 
After that, the proposed Flow-based User Emotion Controlling (FUEC) is used to synthesize waveform by flow matching prediction network conditioned on acoustics prior.  
In this process, the FUEC determines the gradient direction and guidance scale based on the user's emotion instructions by the positive and negative guidance mechanism, which focuses on amplifying the desired emotion while suppressing others. 
Extensive experimental results demonstrate favorable performance compared to several state-of-the-art methods. 
The code and trained models will be made available at  {\href{https://github.com/GalaxyCong/DubFlow}{\textcolor{red}{https://github.com/GalaxyCong/DubFlow}}}. 

\end{abstract}    
\section{Introduction}
\label{sec:intro}

\begin{figure}[t]
    \centering
    \includegraphics[width=1.0\linewidth]{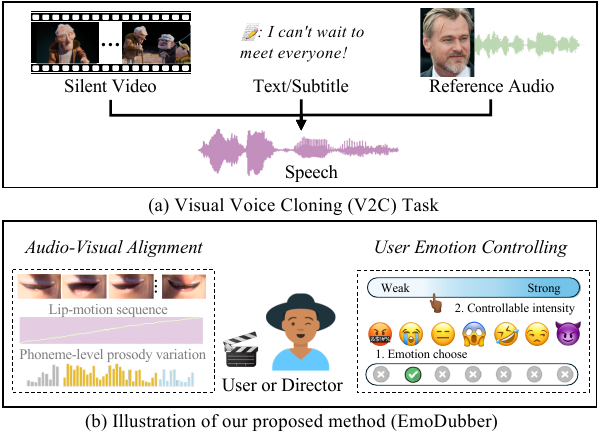}
    \caption{(a) Illustration of the V2C tasks.
    (b) EmoDubber can help users achieve audio-visual sync and maintain clear pronunciation (left), while controlling the intensity of emotions according to the user's intentions (right).   
    }
    \label{fig_intro}
\end{figure}

Movie Dubbing, also known as Visual Voice Cloning (V2C), aims to convert a script into speech with the voice characteristics specified by the reference audio, while aligning lip-sync with the silent video (see Figure~\ref{fig_intro} (a)). 
V2C is far more challenging than conventional text-to-speech (TTS), but has a host of applications, not least in repurposing the vast volumes of existing video to be reproduced by movie creators or enthusiasts. 

Existing dubbing methods broadly fall into two groups.
The first focuses primarily on learning and applying effective speaker style representations. 
For example, V2C-Net~\cite{chen2022v2c} and VDTTS~\cite{hassid2022more} utilize a pre-trained GE2E~\cite{GE2EICASSP} to obtain unique and normalized utterance embedding. 
StyleDubber~\cite{cong2023learning} applies multi-scale learning to apply style cues to the embeddings extracted by GE2E, while Speak2Dub~\cite{zhang2024speaker} adopts pre-training strategy on TTS corpus to improve expressiveness. 
Because these methods rely on simple MSE-based loss~\cite{zhang2024speaker, chen2022v2c} and overall scaling~\cite{cong2023learning}, their capacity to establish accurate correspondence between lip motion and speech is limited. 
The second group of methods aim to generate proper prosody by incorporating visual information from the provided video~\cite{hu2021neural, lee2023imaginary, zhao2024mcdubber, cong2023learning}. 
For example, HPMDubbing~\cite{cong2023learning} applies visual information to speech prosody through hierarchical modeling of lips, face, and scene.  
Recently, MCDubber~\cite{zhao2024mcdubber} operates at the level of context sequence with previous and following facial information to enhance generated speech prosody.  
These methods work at the video-frame and mel-spectrogram levels, however, which means they ignore the role of phoneme-level pronunciations, and thus often generate seemingly mumbled articulation.

Except for the above shorts, most current dubbing models suffer from rigid emotional expression due to lacking controllability.  
Some studies~\cite{iturregui2021audio} suggest that the emotional intensity in dubbing can affect the listener's emotions and psychological perception of the film.
Since the film's post-production can make up for the deficiencies of previous recordings, especially in emotional expression, the actors need to re-record in studio according to the director's instructions until they meet the requirements. %
Unlike previous methods, our model not only satisfies the basic function (lip-sync and clear pronunciation), but also learns to control the attribute and intensity of emotions to meet customized needs, as shown in Figure~\ref{fig_intro} (b). %

In this paper, we propose a novel movie dubbing architecture named EmoDubber, which achieves emotion synthesis with controllable intensity while maintaining audio-visual alignment and clear pronunciation. 
Specifically, we first design a Lip-related Prosody Aligning (LPA) module that controls speech speed by duration level contrastive learning between lip motion and phoneme prosody sequence, which helps the model to reason the correct audio-visual aligning. 
Second, we propose the Pronunciation Enhancing (PE) strategy, which focuses on expanding video-level phoneme sequences by monotonic alignment search and fusing it with the output of LPA by an efficient conformer to improve pronunciation. 
Next, the speaker identity adapting module is used to absorbs the fused sequence from PE and injects style embedding from the reference speaker to target acoustics prior information. 
Finally, the proposed Flow-based User Emotion Controlling (FUEC) aims to generate waveform by optimal-transport conditional flow matching based on acoustic priors while rendering the user-specified emotion.   
It is worth noting that we propose positive and negative guidance mechanisms (PNGM) in FUEC to allow user control emotional intensity flexibly, which 
determines the gradient direction of emotion generation and adjusts the dual guidance scale based on the user’s emotion prompt, amplifying the target emotion and suppressing others.

The contributions of this paper are summarized below:
\begin{itemize}
    \item  We propose EmoDubber, a controllable emotion dubbing architecture to help users specify the emotion they need 
    while satisfying high-quality lip sync and pronunciation. %
    \item  We design a FUEC with positive and negative guidance to dynamically adjust the flow-matching vector field prediction process to achieve intensity control flexibly. %
    \item  We simultaneously achieve high-quality lip sync and clear pronunciation by aligning duration-level contrastive learning and phoneme-enhancing strategy. 
    \item Extensive experimental results demonstrate the proposed Emodubber performs favourablly against state-of-the-art models on three benchmark datasets. %
\end{itemize}

\vspace{-3pt}
\section{Related Work}
\label{sec:formatting}

\vspace{-2pt}
\subsection{Visual Voice Cloning}
\vspace{-2pt}
With the rapid development of deep learning~\cite{tu2023self, tu2024distractors, transcpTang, BeichenInductive, YunbinSMART, XuejingEntity, ye2022unsupervised, TianYun, JunxiPrompt, cui2024stochastic, ZhedongGenerating} and speech synthesis~\cite{ wang2024maskgct, du2024cosyvoice, ye2025emotional}, movie dubbing (\ie, V2C) has advanced in recent years. 
It needs to generate a waveform representing how a text might be said and emphasize to keep in step with the lip motion portrayed by a character. %
Some works focus on improving speaker identity to handle multi-speaker scenes~\cite{chen2022v2c, cong2024styledubber, zhang2024speaker,lu2022visualtts, hassid2022more}.  
For example, 
StyleDubber~\cite{cong2024styledubber} propose multi-scale style adaptor with phoneme and utterance level to strengthen speaker's characteristics. 
Besides, some works attempt to combine visual representation to enhance prosody expressiver~\cite{hu2021neural, lee2023imaginary, cong2023learning, zhao2024mcdubber, zhang2025produbber, zhao2025towards}. 
For example, HPMDubbing~\cite{cong2023learning} is a hierarchical dubbing method by bridging acoustic details with visual information: lip motion, face region, and scene. 
Although the speaker identity and prosody modeling have received attention, existing works still suffer from poor lip-sync and lifeless emotional expression, which is unacceptable in dubbing. 
In this work, we propose EmoDubber, a controllable emotional dubbing architecture to help users specify emotion they need in video, while bringing high quality lip-sync and pronunciation.

\begin{figure*}[!htbp]
    \centering
    \includegraphics[width=1.0\linewidth]{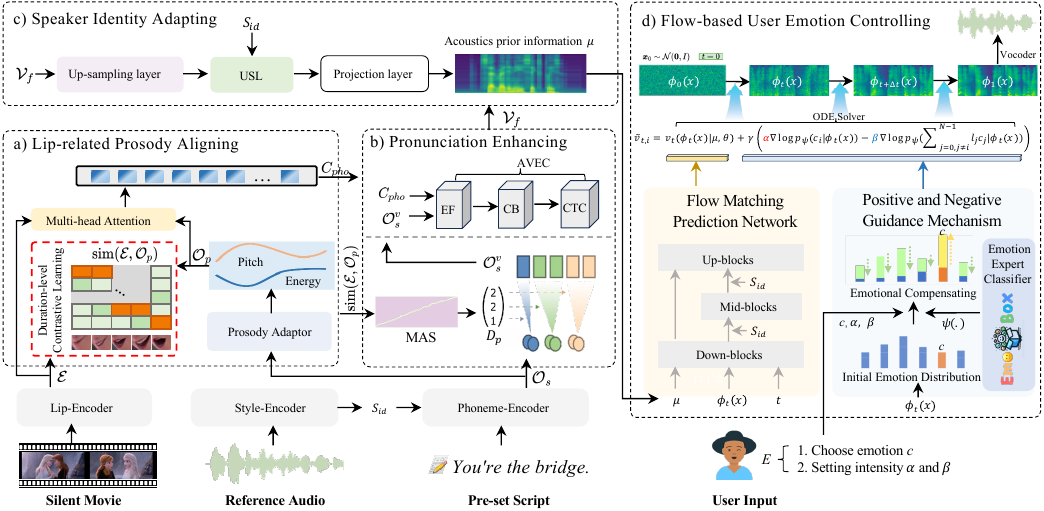}
    \vspace{-0.55cm}
    \caption{
    Architecture of the proposed EmoDubber, which consists of four main components: Lip-related Prosody Aligning (LPA) focuses on learning inherent consistency between lip motion and phoneme prosody by duration level contrastive learning; 
    Pronunciation Enhancing (PE) fuses the output of LPA with expanding phoneme sequence by efficient conformer; 
    Speaker Identity Adapting (SIA) aims to generate acoustics prior information $\mu$ while injecting speaker style; and Flow-based User Emotion Controlling (FUEC) renders user-specified emotion and intensity $E$ in the flow-matching prediction process using positive and negative guidance.  
    }
    \vspace{-13pt}
    \label{fig_archi}
\end{figure*}

\vspace{-2pt}
\subsection{Flow Matching and Classifier Guidance} 
\vspace{-2pt}

Flow Matching~\cite{lipman2022flow} is a simulation-free method to train Continuous Normalizing Flows (CNFs)~\cite{chen2018neural} models, which model arbitrary probability path and capture the probability trajectories represented by diffusion processes~\cite{song2021maximum}. 
It has demonstrated exceptional performance in image generation and geometric domains, such as Stable Diffusion 3~\cite{esser2024scaling}, Lumina-T2X~\cite{gao2024lumina}, and EQUIFM~\cite{YuxuanEquivariant}. 
Due to its advantages of high sampling speed and generation quality, flow matching has attracted significant attention in audio generation~\cite{machaTTS, YiweiVoiceFlow,SungwonFlow}. 
Recently, Matcha-TTS~\cite{machaTTS} introduces optimal-transport conditional flow matching (OT-CFM) for training, which yields an ODE-based decoder to improve the mel-spectrograms fidelity.  
However, these works are limited in the field of TTS and cannot be applied to V2C task.  
The Classifier Guidance~\cite{dhariwal2021diffusion} (CG) has been widely adopted for controlling specific attributes, \eg, text-to-image and emotional TTS~\cite{tang2023emomix, li2024mm, guo2023emodiff}. 
However, the existing emotional TTS using CG only enhances the needed emotion, which struggles to control complex speech containing mixed emotions. 
In this work, we introduce a flow-based user emotion controlling with positive and negative guidance mechanisms, which allows users to manipulate desired emotions and intensity more freely and promotes the development of artificial intelligence in movie dubbing.

\vspace{-5pt}
\section{Proposed Method}
\vspace{-2pt}

\subsection{Overview}
\vspace{-2pt}
Given a silent video clip $V_l$, a reference audio $R_a$, a piece of text $T_p$, and user emotion guidance $E$, 
the goal of EmoDubber is to generate an audio clip $\hat{Y}$ that ensures precise lip-sync and clear pronunciation, while allowing users to control the intensity of emotion by adjusting $E$:
\begin{equation}
    \hat{{Y}} = \mathrm{EmoDubber}(R_a, T_p, V_l, E),  
\end{equation} 
specifically, $E=\{c,\alpha,\beta\}$, where $c$, $\alpha$, and $\beta$ are emotion label, positive weight, and negative weight, respectively. 
The main architecture of the proposed model is shown in Figure~\ref{fig_archi}. 
First, the Lip-related Prosody Aligning (LPA) module absorbs the input information (\ie, $T_p$, $R_a$, and $V_l$) to generate the lip-prosody context sequences with consistent duration cues, like basic pause and speech rate, which are guided by the proposed duration level contrastive learning. 
Next, the Pronunciation Enhancing (PE) strategy focuses on expanding phoneme sequences to video level by monotonic alignment search (MAS) and fusing it with the output of LPA by efficient conformer. 
Then, the Speaker Identity Adapting (SIA) module further decodes the output of the PE to acoustics prior information while introducing style from speaker identity.  
Finally, the Flow-based User Emotion Controlling (FUEC) renders emotions in the flow-matching vector field prediction process using Positive and Negative Guidance Mechanisms (PNGM) according to the user's instructions $E$ to generate desired emotional speech.  

\vspace{-2pt}
\subsection{Lip-related Prosody Aligning}
\vspace{-2pt} 
The core of Lip-related Prosody Aligning (LPA) is  duration-level contrastive learning (DLCL), which aims to capture the correct relevance between prosody and lip motion to bring implicit modeling while focusing on supplying reasonable explicit duration modeling for pronunciation enhancing (PE) strategy. 

\smallskip \noindent\textbf{Extracting Phoneme-level Prosody and Lip-motion Embedding}.  
Firstly, the open-source grapheme-to-phoneme (G2P) is used to obtain the textual phoneme sequence from raw scripts. 
Then, the phoneme encoder with affine transform~\cite{cong2024styledubber} is used to extract style phoneme embeddings $\mathcal{O}_s$: 
\begin{equation}
    \begin{aligned}
        \mathcal{O}_s= \mathrm{PhoEncoder}(T_r{\in \mathbb{R}^{{P}}}, S_{id}),
     \end{aligned}
\end{equation} 
where $\mathcal{O}_s \in \mathbb{R}^{{P}\times{d_{m}}}$, and the ${d_{m}}$ and ${P}$ represent the hidden dimension and length of the phoneme sequence, respectively. 
$S_{id}$ is speaker style embedding, which is extracted by pre-trained speaker encoder from reference audio $R_{a}$, following~\cite{cong2024styledubber, zhang2024speaker}. 
Next, we adopt prosody adaptor~\cite{cong2023learning, zhang2024speaker} to generate phoneme-level prosody variations: pitch embedding ${P}_{pho}$ and energy embedding ${E}_{pho}$. 
Thus, the phoneme-level prosody sequences $\mathcal{O}_{p} = \mathcal{O}_s \oplus {P}_{pho} \oplus {E}_{pho}$ is calculated by combining style phoneme embedding and prosody variations.  
To obtain the lip-motion embedding from the input silent video clip $V_l$, we adopt the same extracting pipeline as~\cite{cong2023learning}: %
\begin{equation}
    \begin{aligned}
        \mathcal{E}= \mathrm{LipEncoder}(M_{roi}\in\mathbb{R}^{{F}\times{D_w}\times{D_h}\times{D_c}}),
     \end{aligned}
\end{equation}
where $M_{roi}$ indicates the mouth Region of Interest (ROI) frame sequence cropped by face landmarks from $V_l$, following~\cite{cong2023learning}.  
$D_w$, $D_h$, and $D_c$ indicate the number of width, height, and channels of images in the mouth ROI frame. 
$F$ denotes the total length of mouth ROI frame. 
$\mathcal{E}\in\mathbb{R}^{{F}\times{d_{m}}}$ denotes the output lip motion embedding. 

\smallskip \noindent\textbf{Duration level Contrastive Learning}.  
Inspired by~\cite{cong2024styledubber,cong2023learning}, we use multi-head attention to capture lip-prosody context sequences by serving lip motion embedding $\mathcal{E}$ as Query and prosody phonemes embedding $\mathcal{O}_{p}$ as Key and Value: 
\begin{equation}
    C_{pho} 
    =  \mathrm{softmax}(\frac{{\mathcal{E}}^\top {\mathcal{O}_{p}}}{\sqrt{d_{m}}}){\mathcal{O}_{p}}^\top,
\label{eq:softAlign}
\end{equation}
where $C_{pho}\in\mathbb{R}^{{{F}}\times{d_{m}}}$ denotes the lip-prosody context sequences with the same length with video clips. 
Instead of non-constraint~\cite{cong2023learning,MinsuDeepVisualForcedAlignment} or simple diagonal-constraint~\cite{cong2024styledubber,hu2021neural}, our LPA module focuses on duration-level contrastive learning (DLCL) to achieve corrected alignment to guarantee monotonicity and subjectivity of weight matrix between prosody and lip-motion sequences: 
\begin{equation}
    \mathcal{L}_{cl}=-\log\frac{\sum \exp\left((\mathrm{sim}^{+}(\mathcal{E}, \mathcal{O}_{p}))/\tau\right)}{\sum \exp((\mathrm{sim}(\mathcal{E}, \mathcal{O}_{p})))}, 
\end{equation} 
where $\mathrm{sim}(\mathcal{E}, \mathcal{O}_{p})$ indicates the attention weight matrix between  $\mathcal{O}_{p}$ and $\mathcal{E}$. 
The positive pair $\mathrm{sim}^{+}(\mathcal{E}, \mathcal{O}_{p})$ is calculated by multiplying $\mathrm{sim}(\mathcal{E}, \mathcal{O}_{p})$ with correct duration-level correspondence $M_{lip, pho}^{gt}$:   
\begin{equation}
    \mathrm{sim}^{+}(\mathcal{E}, \mathcal{O}_{p})=\mathrm{sim}(\mathcal{E}, \mathcal{O}_{p})\times{M_{lip, pho}^{gt}}, 
\end{equation} 
where $M_{lip, pho}^{gt}$ is a ``0-1'' matrix with $P$-th row and $F$-th column and satisfies the monotonicity and surjectivity.  
The value ``1'' represents correct correspondence between textual phoneme and lip motion through Montreal Forced Aligner (MFA)~\cite{mcauliffe2017montreal} model and coefficient between Frames per Second of the video (FPS) and sampling rate (SR).  
In this case, the DLCL encourages the positive pair to have a higher similarity to ensure the phoneme prosody unit focuses on the strongly related part in lip motion sequence.

\vspace{-2pt}
\subsection{Pronunciation Enhancing}
\vspace{-2pt}
The proposed Pronunciation Enhancing (PE) strategy aims to generate video-level phoneme enhancement sequences and fuse it with lip-prosody context sequences.

\noindent\textbf{Explicit duration based Expanding}. 
To obtain the duration of each phoneme unit directly from the learnable attention weight matrix $\mathrm{sim}(\mathcal{E}, \mathcal{O}_{p})$, we use monotonic alignment search (MAS)~\cite{JaehyeonGlow} for explicit alignment. 
Specifically, MAS implements dynamic programming algorithms to find the optimal alignment path on matrix $\mathrm{sim}(\mathcal{E}, \mathcal{O}_{p})$: 
\begin{equation} \mathcal{O}_s^v=\mathrm{LR}(D_p, \mathcal{O}_s), 
\label{eq:hardAlign}
\end{equation} 
specifically, $D_p=\mathrm{MAS}(\mathcal{E}, \mathcal{O}_{p})\in\mathbb{R}^{{P}\times{1}}$ denotes the explicit duration for each phoneme unit, which records the integer multiple alignment between phonemes and video frames.  
The $\mathrm{LR}{(\cdot})$ is a length regulator, which aims to expand style phoneme sequences $\mathcal{O}_s\in\mathbb{R}^{{P}\times{d_{m}}}$ to video-level phoneme enhancement sequences $\mathcal{O}_s^v\in\mathbb{R}^{{F}\times{d_{m}}}$.

\noindent\textbf{Efficient Conformer based Fusing}. 
Inspired by~\cite{MaximeEfficient}, we use audio-visual efficient conformer (AVEC) to model both local and global dependencies using convolution and attention to reach better fusing performance on two kinds of feature: lip-prosody context sequences
 $C_{pho}$ (from Eq.~\ref{eq:softAlign}) and phoneme enhancement sequences $\mathcal{O}_s^v$ (from Eq.~\ref{eq:hardAlign}): 
\begin{equation} 
\mathcal{V}_{f} =\mathrm{Conformer}(C_{pho}, \mathcal{O}_s^v), 
\label{eq:fused}
\end{equation}
where $\mathcal{V}_{f}\in \mathbb{R}^{{F}\times{d_{m}}}$ indicates the fused intermediate feature. 
The $\mathrm{Conformer}(\cdot)$ represents AVEC, which consists of an early fusion (EF) strategy~\cite{PingchuanConformers} to reduce model complexity, 5 Conformer blocks (CB) without downsampling, and connectionist temporal classification (CTC)~\cite{GravesConnectionist, MinsuLiptoSpeech} layer to maximize the sum of probabilities of correct target phoneme to ensure pronunciation.

\vspace{-2pt}
\subsection{Speaker Identity Adapting}
\vspace{-2pt}

The Speaker Identity Adapting (SIA) aims to generate acoustics prior information $\mu$ in mel-spectrogram level from $\mathcal{V}_{f}$ (Eq.~\ref{eq:fused}) and speaker style embedding $S_{id}$:   
\begin{equation} 
\mu =\mathrm{Proj}(\mathrm{USL}(\mathrm{Up}(\mathcal{V}_{f}), S_{id})),
\label{eq:mu_cond}
\end{equation}
where $\mu \in\mathbb{R}^{{M_l}\times{d_a}}$, and the ${M_l}$ and $d_a$ represent the length and hidden dimension of desired mel-spectrogram sequence. 
The SIA consists of up-sampling layer, utterance-level style learning (USL) module~\cite{cong2024styledubber}, and projection layer~\cite{cong2024styledubber}. 
Firstly, we upsample the time axis of  $\mathcal{V}_{f}$ to mel-spectrogram level by applying two layers of 2D convolutions~\cite{YochaiLipVoicer}. 
Then, we use USL to inject the style information from style embedding $S_{id}$.  
Finally, the projection layer is used to project the output feature to target dimension.

\vspace{-2pt}
\subsection{Flow-based User Emotion Controlling} 
\vspace{-2pt}

For given user's emotion instructions $E=\{c,\alpha,\beta\}$, the proposed Flow-based User Emotion Controlling (FUEC) focuses on determining the gradient direction and guidance scale based on $E$ while iteratively converting noise into the mel-spectrogram to inject emotion expressiveness.  
It consists of Flow Matching Prediction Network (FMPN) and Positive and Negative Guidance Mechanisms (PNGM).

\noindent\textbf{Flow Matching Prediction Network}. 
Given the mel-spectrograms data space with data point $M$, where $M \sim q(M)$ and $q(M)$ is an unknown data distribution of mel-spectrograms, a possible approach to sample $M$ from $q(M)$ is to give a probability density path defined as $p_t(x)$ where $t\in [0,1]$, $p_0(x)=\mathcal{N}(x;\boldsymbol{0},\boldsymbol{I})$ and $p_1(x)\approx q(x)$. Flow matching model can estimate the probability density path, gradually transforming noise $x_0\sim p_0(x)$ into mel spectrogram $M\sim q(M)$. Here, we train Flow Matching Prediction Network (FMPN) based on optimal-transport conditional flow matching (OT-CFM) with a linear interpolation flow $\phi_t(x)=(1-(1-\sigma_{\min})t)x_0+tM$, gradually transform noise $x_0$ to mel-spectrogram M from $t=0$ to $t=1$. Its gradient vector field is $u_t(\phi_t(x)|M)=M-(1-\sigma_{min})x_0$, facilitating fast training and inference from noise to mel spectrogram due to its linear and time-invariant properties.
The training objective is to train FMPN donated by $\theta$ to predict the gradient vector field of $\phi_t(x)$:
\begin{equation}
\resizebox{1.0\linewidth}{!}{
$
\mathcal{L_\theta}=\mathbb{E}_{t,q(M),p_t(x|\mu,M)} ||v_t(\phi_t(x)|\mu,\theta)-u_t(\phi_t(x)|M)||^2,
$
}\label{eqflow}
\end{equation}
where $v_t(\phi_t(x)|\mu,\theta)$ is the predicted gradient vector field of $\phi_t(x)$ according to the acoustics prior information $\mu$. Then, we can solve the ODE $d\phi_t(x)=v_t(\phi_t(x)|\mu,\theta)dt$ from $t=0$ to $t=1$ to generate the target mel-spectrogram $\hat{M}$ from noise $x_0$. 
Besides, we use the style affine learning in FMPN to adapt multi-speaker scenarios effectively.

\begin{table*}[!t]
  \centering
  \resizebox{1.0\linewidth}{!}
  {
    \begin{tabular}{c|cccccc|cccccc}
    \hline
    Setting & \multicolumn{6}{c|}{Dubbing Setting 1.0} & \multicolumn{6}{c}{Dubbing Setting 2.0} \\ 
    \toprule
    Methods
    & LSE-C $\uparrow$ 
    & LSE-D $\downarrow$   
    & WER $\downarrow$ 
    & SECS $\uparrow$
    & MCD $\downarrow$
    & MCD-SL $\downarrow$
    & LSE-C  $\uparrow$ 
    & LSE-D  $\downarrow$  
    & WER $\downarrow$ 
    & SECS $\uparrow$ 
    & MCD $\downarrow$
    & MCD-SL $\downarrow$
    \\ 
    \midrule
    GT & 8.12 & 6.59  &  3.85 & 100.00 & 0.0  & 0.0 & 8.12 & 6.59 & 3.85 & 100.00 & 0.0 & 0.0 \\
    \midrule
    Fastspeech2*~\cite{ren2020fastspeech} & 3.34 & 11.60 & 15.33 & 79.26 &  5.88 & 6.73 & 3.34 & 11.60 & 15.33 & 79.26 & 5.88  & 6.73 \\ 
    StyleSpeech*~\cite{Dongchan2021StyleSpeech} & 2.06 & 12.27 & 79.14 & 63.00 & 7.64  & 9.87 & 2.12 & 12.14 & 80.01 & 60.06 & 8.31  & 10.28  \\
    Face-TTS~\cite{lee2023imaginary} & 1.98 & 12.50 & 62.24 & 59.56 &  7.51 & 12.59 & 1.96 & 12.53 & 68.13 & 53.44 & 7.64  & 12.79  \\
    V2C-Net~\cite{chen2022v2c} & 1.97 & 12.17 & 90.47 & 51.52 & 6.25  & 8.31 & 1.82 & 12.09 & 94.59 & 44.19  & 6.74  &  9.04 \\
    HPMDubbing~\cite{cong2023learning} & 7.85 & 7.19 & 16.05 & 85.09 & 6.12  & 7.25 & 3.98 & 9.50 & 29.82 & 73.55 & 6.91 & 8.56  \\
    Speaker2Dub~\cite{zhang2024speaker} & 3.76 & 10.56 & 16.98 & 74.73 & 7.67  & 7.89  & 3.45 & 11.17 & 18.10 & 69.28 & 8.06 & 8.21 \\
    StyleDubber~\cite{cong2024styledubber} & 3.87 & 10.92  & 13.14 & 87.72  & \textbf{5.41}  & \textbf{5.73} & 3.74 & 11.00 & 14.18 & 82.07  &  \textbf{6.01} &  \textbf{6.36} \\
    \midrule
     Ours & \textbf{8.11}  &  \textbf{6.92}    &  \textbf{11.72}      &  \textbf{90.62}   &  {5.87}  &  {5.87}  &  \textbf{8.09}   & \textbf{6.96} & \textbf{12.81}  &\textbf{85.06}  & {6.51} & {6.51}\\
    \bottomrule
    \end{tabular}
    }
    \vspace{-0.35cm}
    \caption{Results on Chem benchmark. The method with “*” refers to a variant taking video embedding as an additional input following. 
    }
  \label{result_Chem}%
  \vspace{-0.4cm}
\end{table*}%

\begin{table}[!tbp]
  \centering
  \resizebox{1.0\linewidth}{!}
  {
    \begin{tabular}{lccccccc}
    \toprule
    Methods & LSE-C $\uparrow$ & LSE-D $\downarrow$ & WER $\downarrow$ & MOS-S $\uparrow$ & MOS-N $\uparrow$ \\
    \midrule
    StyleDubber~\cite{cong2024styledubber} & 6.17  &  9.11 
     &  15.10 & 4.03$\pm$0.10  & 3.85$\pm$0.15 \\
    Speaker2Dub~\cite{zhang2024speaker} & 4.83 & 10.39
     & 15.91 & 3.98$\pm$0.09 &  4.01$\pm$0.13\\
     \midrule 
    Ours  & \textbf{7.40} & \textbf{6.65} & \textbf{14.03} & \textbf{4.07$\pm$0.09} & \textbf{4.05$\pm$0.06} \\
    \bottomrule
    \end{tabular}%
    } 
    \vspace{-0.35cm}
  \caption{The zero shot results under Dub 3.0 setting, which use unseen speaker as refernce audio.} 
  \vspace{-0.3cm}
  \label{tab_setting3_Zero_shot}%
\end{table}

\noindent\textbf{Positive and Negative Guidance Mechanisms.} 
Inspired by the emotions in human speech that are often blended rather than single~\cite{KunZhouMixed2023}, where multiple emotions can naturally overlap or co-exist, we propose Positive and Negative Guidance Mechanisms (PNGM) based on Classifier Guidance~\cite{dhariwal2021diffusion} to guide $v_t(\phi_t(x)|\mu,\theta)$ with specific emotion $c$.  
Unlike previous emotion synthesis methods~\cite{KunZhouMixed2023, guo2023emodiff}, PNGM allows users to perform more flexible manipulations in flow-matching  prediction process by introducing positive and negative guidance to enhance the desired emotion and suppress others.

Suppose we have a well-trained FMPN $\theta$ predicting $v_t(\phi_t(x)|\mu,\theta)$ and an emotional expert classifier $\psi$ which can predict the probability $p_\psi(c|\phi_t(x))$ that $\phi_t(x)$ belongs to emotion $c$. 
Note that our emotional expert classifier is pre-trained on multiple large-scale emotion datasets recorded in Emobox~\cite{ma2024emobox}, enabling a better determination of real human emotions and improving the performance of FMPN-guided emotion. 
We can set emotion classes as $\{c_0,\cdots,c_{N-1} \}$ one-hot vector with $N$ kinds of emotions on mel-spectrogram data $M$. 
The emotional softmax logit of $\phi_t(x)$ predicted by the emotional classifier $\psi$ is $l_\psi(\phi_t(x))=[l_0,\cdots,l_{N-1}]$, which can be seen as the emotional mix ratio of $\phi_t(x)$, and the mix emotion is $c_M=\sum_{i=0}^{N-1}l_ic_i$. 
To enhance emotion $c_i,i\in \{0,\cdots,N-1\}$, we can use positive guidance to guide $M$ toward the direction of $c_i$ and use negative guidance to suppress others, which can be formulated as:
\begin{equation}
\resizebox{1\linewidth}{!}{
$
    \begin{aligned}
\tilde{v}_{t,i}&=v_t(\phi_t(x)|\mu,\theta) \\ &+ 
\gamma \Big(\alpha \nabla \log p_{\psi}(c_i|\phi_t(x))- \beta \nabla \log p_{\psi}(\sum_{j=0,j\neq i}^{j=N-1}l_jc_j|\phi_t(x))  \Big),
    \end{aligned}
$
}
\label{eq:guidance}
\end{equation}
where $\gamma$ controls the total degree of PNGM, $\alpha$ is the positive guidance scale controlling the emotion of $c_i$ and $\beta$ is the negative guidance scale controlling the degree weakening other emotions. 
This way, users can change $\alpha$ and $\beta$ to control the emotion expressiveness of synthesized $\hat{{M}}$. 
Finally, the generated emotional mel-spectrograms $\hat{{M}}$ are converted to time-domain wave $\hat{{Y}}$ via the powerful vocoder.

\section{Experimental Results}
\subsection{Implementation Details}
Video frames are sampled at 25 FPS and all audios are resampled to 16kHz. 
The lip region is resized to 96 $\times$ 96 and pre-trained on ResNet-18, following~\cite{martinez2020lipreading, ma2020towards}. 
The window length, frame size, and hop length in STFT are
640, 1,024, and 160, respectively.  
Each block had one Transformer layer with hidden dimensionality 256 and style affine layers~\cite{cong2024styledubber}, 2 heads, attention dimensionality 64, and snakebeta~\cite{SangBigVGAN} activations. 
We use 8 heads for multi-head attention in LPA with 256 hidden sizes. 
The temperature coefficient $\tau$ of $\mathcal{L}_{cl}$ as 0.1. 
There are 5 conformer blocks in AVEC. 
We trained V2C, Chem, and GRID with batch sizes 64, 32, and 64, respectively. 
The Emotion Expert Classifier  $\psi$ is trained on 13 emotional datasets with more than 50,000 emotional audio recordings in Emobox~\cite{ma2024emobox}, which collects large-scale Speech Emotion Recognition (SER) benchmarks. 
The structure of the classifier is a 6-layer 1D CNN with BatchNorm, LeakyReLU, and Dropout. 
Only phoneme encoder, USL, and flow decoder are pre-trained on LibriSpeech~\cite{LibrispeechVassil} to fair compare with the pre-trained dub method~\cite{zhang2024speaker}. 
During the inference process, $\gamma$ is set to 15, while $\alpha\in[0,9]$ and $\beta\in[0,2]$. 
Both training and inference are implemented with PyTorch on a GeForce RTX 4090 GPU. 
More details are given in the Appendix.

\subsection{Datasets}

\noindent\textbf{Chem} is a popular dubbing dataset recording a chemistry teacher speaking in the class~\cite{prajwal2020learning}. 
It is collected from YouTube, with a total video length of approximately nine hours. 
For complete dubbing, each video has clip to sentence-level~\cite{hu2021neural}. %
The number of train, validation, and test data are 6,082, 50, and 196, respectively. 

\noindent\textbf{GRID} is a dubbing benchmark for multi-speaker dubbing~\cite{cooke2006audio}.  
The whole dataset has 33 speakers, each with 1,000 short English samples. 
All participants are recorded in studio with unified background. 
The number of train and test data are 32,670 and 3,280, respectively.

\subsection{Evaluation Metrics}

\noindent\textbf{Audio-visual Sync Evaluation}.  
To evaluate the synchronization between the generated speech and the video quantitatively, we adopt Lip Sync Error Distance (LSE-D) and Lip Sync Error Confidence (LSE-C) as our metrics, which are widely used to lip reading~\cite{YochaiLipVoicer}, talking face~\cite{JiadongSeeing,YoungjoonFaces}, and video dubbing task~\cite{hu2021neural,lu2022visualtts}. 
These metrics are based on the pre-trained SyncNet~\cite{chung2016out}, which can explicitly test for lip synchronization in unconstrained videos in the wild~\cite{prajwal2020lip,chung2016out}. 
Compared to the length metric MCD-SL~\cite{chen2022v2c}, we believe that LSE-C and LSE-D can more accurately measure the synchronization of vision and audio.

\noindent\textbf{Speech Quality Evaluation}. The Word Error Rate (WER)~\cite{AndrewWER} is used to measure pronunciation accuracy by using Whisper-V3~\cite{whisper} as the ASR model. 
To evaluate the timbre consistency between the generated dubbing and the reference audio, we employ the speaker encoder cosine similarity (SECS) following~\cite{cong2024styledubber,zhang2024speaker} to compute the similarity of speaker identity. 
Besides, we adopt the Mel Cepstral Distortion Dynamic Time Warping (MCD) and speech length variant (MCD-SL)~\cite{battenberg2020location, muller2007dynamic, chen2022v2c} to measure the difference between generated speech and real speech.

\noindent\textbf{Emotional Evaluation}. We use the Intensity Score, the average softmax logit of the target emotion, which ranges from 0 to 1, to measure the emotional intensity of generated audio. While previous works~\cite{guo2023emodiff, KunZhouMixed2023} use average classification probability, they fail to distinguish varying intensities within the same emotion class. The average softmax logit provides a finer-grained measure, allowing for a more effective evaluation of emotion intensity differences in audio.

\begin{table}[!tbp]
  \centering
  \resizebox{1.0\linewidth}{!}
  {
    \begin{tabular}{lccccccc}
    \toprule
    \# & Methods   & LSE-C $\uparrow$ & LSE-D $\downarrow$ & WER $\downarrow$ & SECS $\uparrow$ & MCD $\downarrow$  \\
    \midrule
    1 & w/o PE   & \textbf{8.10} & \textbf{6.81} & 53.36 & 84.92& 7.19 \\
    2 & w/o SIA & 8.05 & 6.99 & 13.07 & 82.04& 6.72  \\ 
    3 & w/o LPA  & 4.47 & 10.45 & 36.31  & 85.03 &  7.41  \\
    \midrule
    4 & Full model  & {8.09}    &  {6.96}   &  \textbf{12.81} & \textbf{85.06}  & \textbf{6.51}  \\ 
    \bottomrule
    \end{tabular}%
    } 
 \vspace{-0.35cm}
  \caption{Ablation study of the proposed EmoDubber on the Chem benchmark dataset with 2.0 setting.} 
  \vspace{-0.4cm}
  \label{tab_ablation}%
\end{table}

\noindent\textbf{Subjective Evaluation.} To further evaluate the quality of generated speech, we conduct a human study using a subjective evaluation metric, following the settings in~\cite{chen2022v2c}. 
Specifically, we adopt the MOS-naturalness (MOS-N) and MOS-similarity (MOS-S) to assess the naturalness of the generated speech and the recognization of the desired voice.

\begin{table*}[!t]
  \centering
  \resizebox{1.0\linewidth}{!}
  {
    \begin{tabular}{c|cccccc|cccccc}
    \hline
    Setting & \multicolumn{6}{c|}{Dubbing Setting 1.0} & \multicolumn{6}{c}{Dubbing Setting 2.0} \\ 
    \toprule
    Methods
    & LSE-C $\uparrow$ 
    & LSE-D $\downarrow$   
    & WER $\downarrow$ 
    & SECS $\uparrow$
    & MCD $\downarrow$
    & MCD-SL $\downarrow$
    & LSE-C  $\uparrow$ 
    & LSE-D  $\downarrow$  
    & WER $\downarrow$ 
    & SECS $\uparrow$ 
    & MCD $\downarrow$
    & MCD-SL $\downarrow$
    \\   
    \midrule
    GT & 7.13 & 6.78 & 22.41 & 100.00 & 0.00  & 0.00 & 7.134 & 6.786 & 22.41 & 100.00 & 0.00  & 0.00  \\
    \midrule
    Fastspeech2*~\cite{ren2020fastspeech} & 5.01 & 9.79 & 19.61 & 11.35 &   7.24 & 7.95 & 5.01 & 9.79 &  19.61 & 11.35 & 7.24  & 7.95 \\
    StyleSpeech*~\cite{Dongchan2021StyleSpeech} & 5.90 & 9.24 & 22.62 & 90.04 &   5.74 & 5.88 &  4.79 & 10.28 & 19.82 & 59.58 & 7.01 &  7.82  \\
    Zero-shot TTS*~\cite{YixuanZhou} & 5.03 & 10.02 &20.05 & 85.93 &  5.75 & 6.40  & 4.48 & 10.54 & 21.05 & 81.34 &  6.27  &  7.29 \\
    Face-TTS~\cite{lee2023imaginary} & 4.69 & 10.14 & 44.37 & 82.97 &  7.44 & 8.16 & 4.55  & 10.27 & 39.05 & 34.14 & 7.77 & 8.59  \\
    V2C-Net~\cite{chen2022v2c} & 5.59 & 9.52 & 47.82 & 80.98 &  6.79  & 7.23 & 5.34 & 9.76 & 49.09 & 71.51  & 7.29 &  7.86 \\
    HPMDubbing~\cite{cong2023learning} & 5.76 & 9.13 & 45.51 & 85.11 &  6.49  & 6.78 & 5.82 &  9.10 & 44.15 & 71.99 &  6.79  & 7.09  \\
    StyleDubber~\cite{cong2024styledubber} & 6.12 & 9.03 & 18.88 & 93.79 & 5.61  & 5.69 & 6.09 & 9.08  & 19.58 & \textbf{86.67} & 6.33 &  6.42 \\
    Speak2Dub~\cite{zhang2024speaker} & 5.27 & 9.84 & \textbf{17.07} & \textbf{94.50} & 5.34   & 5.45 & 5.19 & 9.93 & \textbf{17.42} & 85.76  & 6.17 &  6.43 \\
    \midrule
     Ours & \textbf{7.12}  &  \textbf{6.82}    &  {18.53}     &  {92.22}  &  \textbf{3.13}  &  \textbf{3.13}  &  \textbf{7.10}  & \textbf{6.89}  &  {19.75} &  {86.02} & \textbf{3.92}  & \textbf{3.92}  \\
    \bottomrule
    \end{tabular}
    }
    \vspace{-6pt}
    \caption{Results on GRID benchmark. The method with “*” refers to a variant taking video embedding as an additional input following. 
    }
  \label{result_Grid}
  \vspace{-10pt}
\end{table*}%

\begin{figure*}
    \centering
    \includegraphics[width=1.0\linewidth]{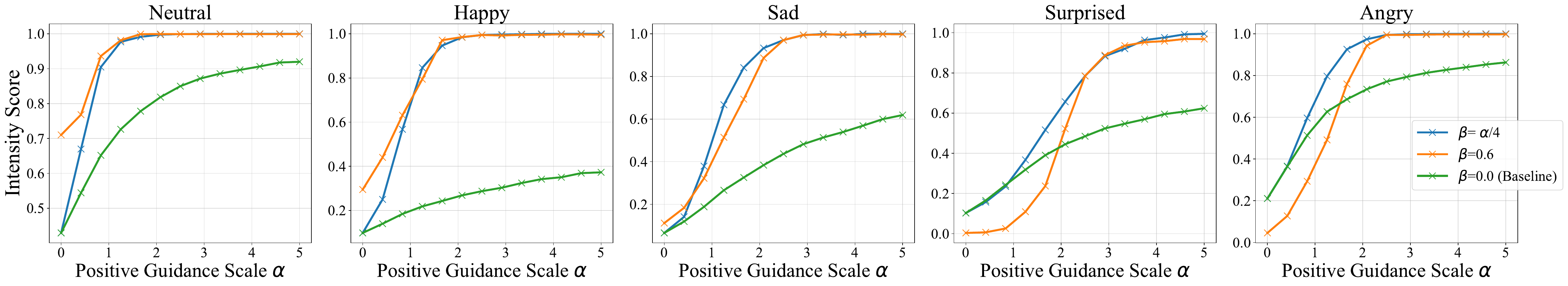}
    \vspace{-14pt}
    \caption{Intensity performance of EmoDubber on Chem. 
    The horizontal axis shows the positive guidance $\alpha$, and vertical axis displays the Intensity Score (IS), with different curves for various negative guidance $\beta$. Higher IS indicate stronger emotional intensity in audio.}
    \vspace{-10pt}
    \label{fig:intensity_ratio}
\end{figure*}

\subsection{Comparison with SOTA Dubbing Methods}
\vspace{1mm} 

To compare with SOTA dubbing model without the function of emotion control, we remove the PNGM, \ie, maintaining Figure~\ref{fig_archi} (a)-(c) and flow matching prediction network to generate waveform. 
We compare with the recent dubbing baselines to comprehensively analyze. 

\noindent\textbf{Results on the Chem Dataset}. 
{
As shown in Table~\ref{result_Chem}, our method achieves the best performance on almost all metrics on Chem benchmark. %
Although StyleDubber can achieve good MCD and MCD-SL, they performed poorly on LSE-D and LSE-C, which reflects that they did not achieve true lip-sync. 
In contrast, our method achieves the best LSE-C and LSE-D, with absolute improvements of 4.24\% and 4.0\%, as well as the best WER with an improvement of 10.80\%, demonstrating the effectiveness of proposed method to maintain high-quality audio-visual alignment and clear pronunciation simultaneously. 
Regarding speaker similarity (see SECS), the proposed method outperforms SOTA baseline StyleDubber with an absolute margin of 2.9\%. 
Please note that setting 2 is more challenging than setting 1, which requires the model to have strong robustness. 
Despite challenging, EmoDubber still has a lead in lip-sync, pronunciation, and speaker identity. 

\vspace{1mm}
\noindent\textbf{Results on the GRID Dataset}.  
We report the GRID result in Table~\ref{result_Grid}. 
Our method is currently the only one that achieves the best performance in terms of both lip-sync (see LSE-C and LSE-D) and pronunciation clarity (see WER), whether in setting or setting 2. 
Specifically, our method significantly improves 19.46\% LSE-C and 24.12\% LSE-D on challenging setting 2, which indicates the effectiveness of the proposed approach in achieving accurate lip sync even in multi-speaker dubbing scenes. 
In addition, our method also achieves competitive results in WER, only slightly lower than the best pre-trained model Speak2Dub. 
But it turns out that our WER result (18.53\%) exceeds the ground truth WER result (22.41\%), which means that the intelligibility has reached the acceptable range for humans. 
Finally, our method achieves lowest MCD and MCD-SL compared to all baselines, which indicates our method achieves minimal acoustic difference in challenging setting 2.0. 

\vspace{1mm}
\noindent\textbf{Results on the Speaker Zero-shot test}. 
This setting uses the audio of unseen characters (from another dataset) as reference audio to measure the generalizability of the dubbing model. 
Here, we use the audio from the Chem dataset as reference audio to measure the GRID dataset. 
Since there is no target audio at this setting, we only compare LSE-C/D and WER, and make subjective evaluations. 
As shown in Table~\ref{tab_setting3_Zero_shot}, We were surprised to find that our method outperforms the SOTA dubbing methods (StyleDubber and Spk2Dub) on all metrics. 
In particular, our method surpasses the current best dubbing method Spk2Dub in WER, which reflects that our model is more robust in maintaining clear pronunciation in unseen speaker scenes. 
Furthermore, the proposed method still maintains the leading position in audio-visual synchronization (see LSE-C and LSE-D), which other SOTA dubbing methods cannot achieve.

\begin{figure*}[t]
    \centering
    \includegraphics[width=1.0\linewidth]{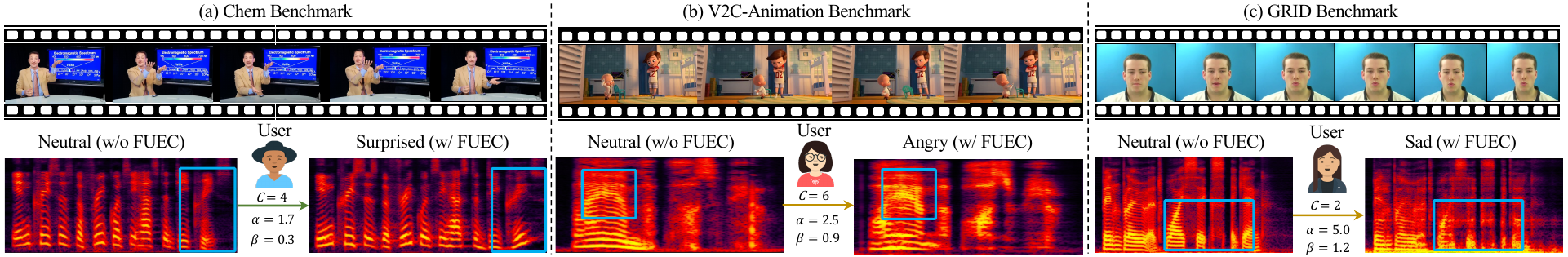}
    \caption{Visualization of audio samples generated by EmoDubber: one uses the proposed FUEC to guide emotions by users, and the other does not (Neutral).  
    The green rectangles highlight key regions that have significant differences in emotional expressiveness. 
    }
    \vspace{-14pt}
    \label{fig_visual}
\end{figure*}
\vspace{1mm}

\begin{figure}
\begin{minipage}[]{0.49\linewidth}
    \centering
    \includegraphics[width=1.0\linewidth]{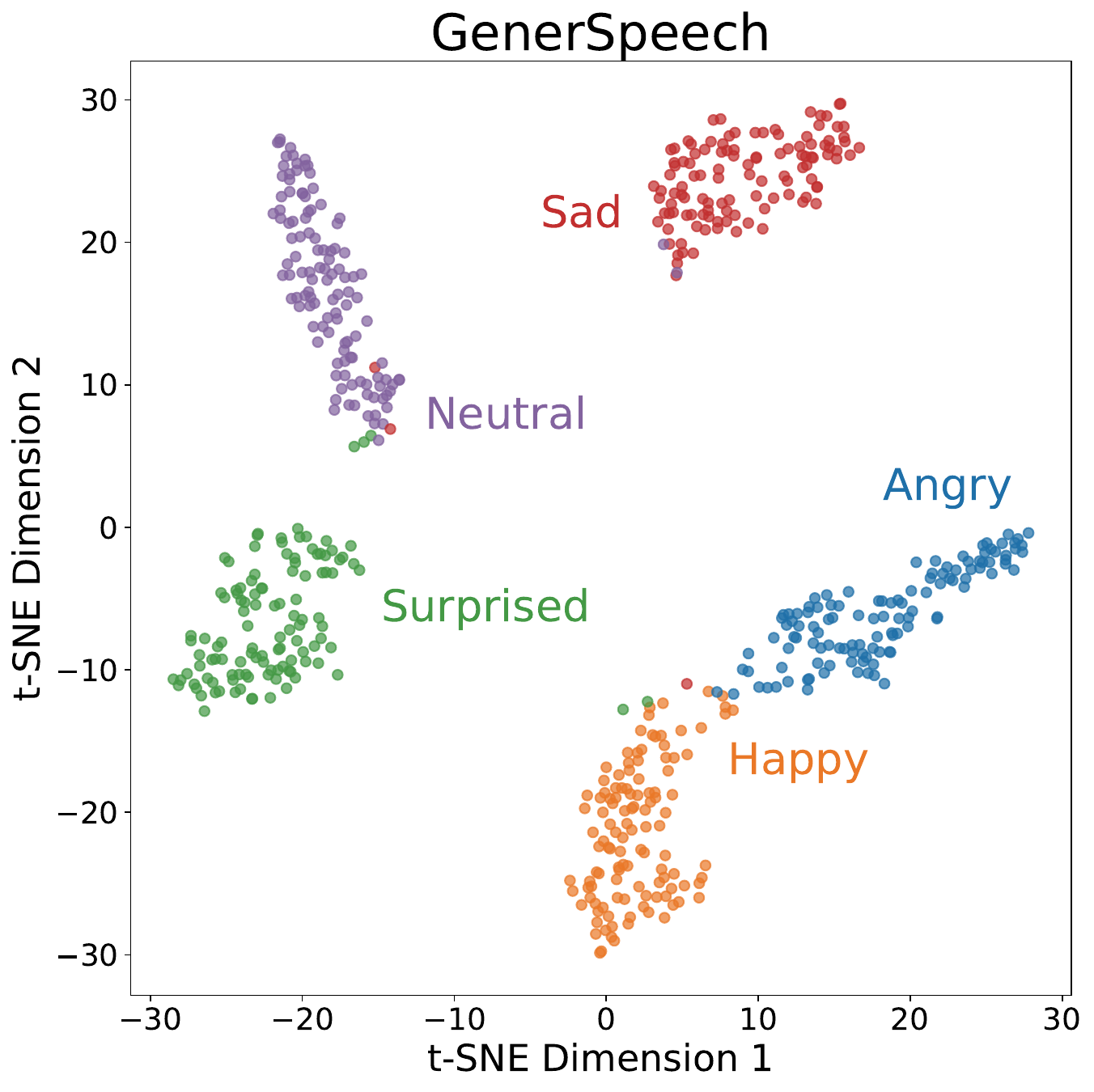}
\end{minipage}
\hfill
\begin{minipage}[]{0.49\linewidth}
    \centering
    \includegraphics[width=1.0\linewidth]{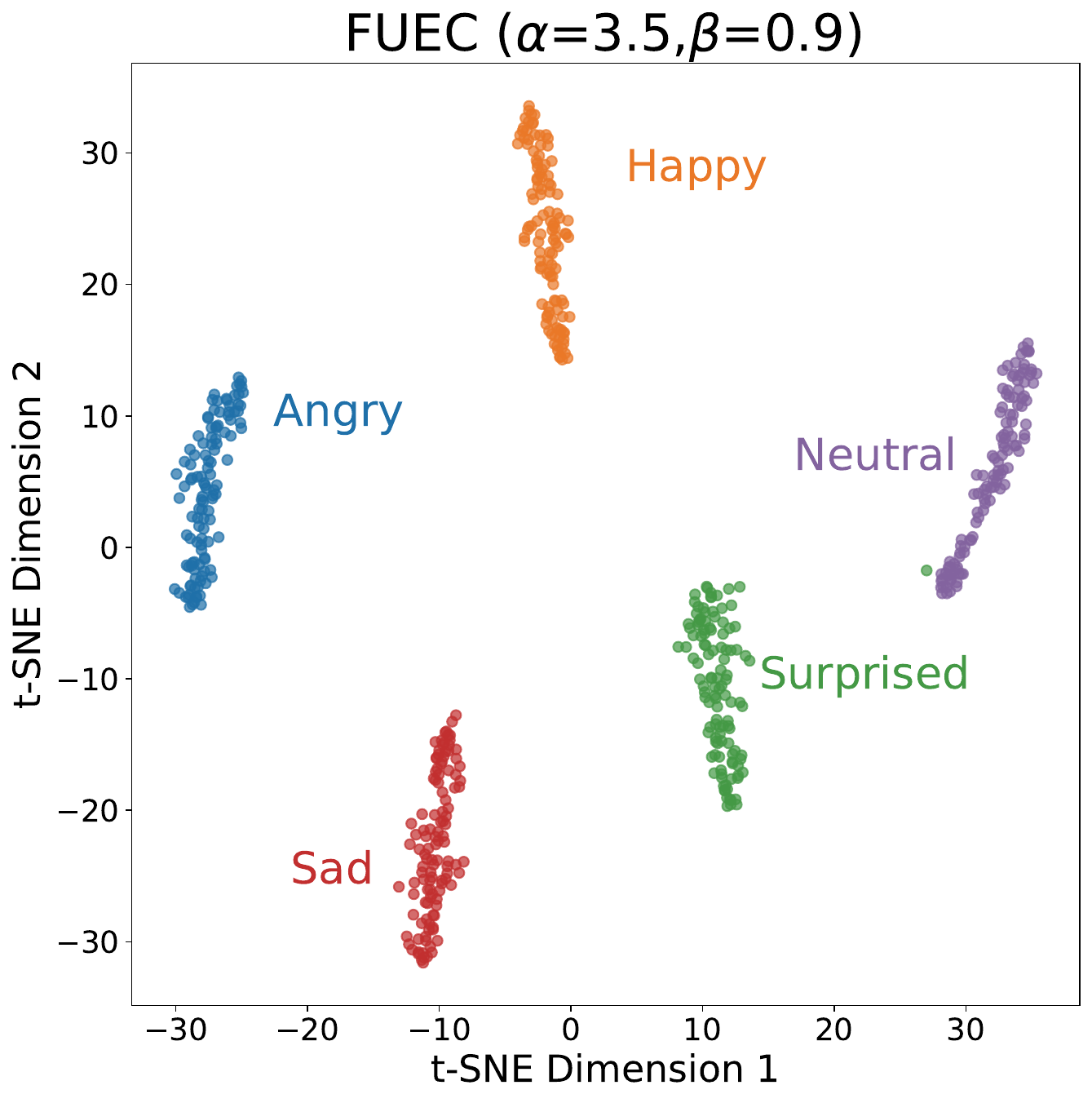}
\end{minipage}
\vspace{-6pt}
\caption{Visual results of emotional audio features by t-SNE, the TTS baseline is shown on the left and EmoDubber on the right. 
}
\vspace{-0.4cm}
\label{fig:tsne_v2c}
\end{figure}

\noindent\textbf{Ablation Studies}. 
The ablation results are presented in Table~\ref{tab_ablation}. 
It shows that all three modules contribute significantly to the overall performance, and each module has a different focus. 
After removing the PE, the WER severely drop. 
This reflects that the PE achieves better pronunciation by fusing video-level phoneme enhancement sequences with explicit duration. 
In contrast, the SECS is most affected by SIA, which indicates decoding mel-spectrograms by introducing global style is beneficial to identity recognition. 
Finally, the performance of LSE-C and LSE-D drops the most when removing LPA. 
It proves the effectiveness of modeling the relevance between lip motion and phoneme prosody by duration-level contrastive learning, which is beneficial to reason correct audio-video aligning.

\vspace{-6pt}
\subsection{Emotional Controlling Evaluation}

\vspace{-6pt}
\noindent\textbf{Intensity Controlling Results}. 
Figure~\ref{fig:intensity_ratio} shows the intensity-controlling results of EmoDubber on the Chem benchmark.  
The higher Intensity Score means a stronger emotional intensity of speech.
We vary $\alpha$ from 0.0 to 5.0 and $\beta$ from 0.0 to 2.0 and present results for two cases: one with a fixed $\beta$ value and another where $\beta$ varies with $\alpha$ to compare with the traditional baseline with no dual-scale guidance. 
Compared to the baseline, EmoDubber offers a wider range of intensity control, evident from our broader Intensity Score range from $\alpha=0.0$ to $\alpha=5.0$. 
Additionally, our method enables stronger emotional modulation, significantly achieving a higher Intensity Score than the baseline when $\alpha=5.0$. 
This demonstrates that our EmoDubber supports a broader range of emotional control and enables the generation of audio with stronger emotional intensity by FUEC. 
We also note that the combination values of $\alpha$ and $\beta$ can be more diverse, supporting a wider range of emotional intensity controls.

\noindent\textbf{Emotion Zero-shot Conversion}. 
To verify the emotion generalization on the emotionless dubbing dataset, we re-synthesis speech with five kinds of emotion on the Chem dataset (no emotion label). 
We use the publicly available GenerSpeech~\cite{RongjieGenerSpeech} as our baseline, which is a SOTA TTS model for emotional transfer in out-of-domain. 
As shown in Figure~\ref{fig:tsne_v2c}, we visualize the audio features using t-SNE~\cite{van2008visualizing}. 
Compared to the TTS baseline, EmoDubber demonstrates better emotion differentiation, highlighting its ability to generate audio with distinct emotions and generalize to other emotionless videos. 
Besides, it is worth noting that our model can free to adjust emotion intensity control in lip-sync dubbing, which is not possible with current emotional TTS baselines.

\noindent\textbf{Emotion Speech Quality Results}. 
To verify speech quality after emotion control, we test 7 kinds of emotional speech with each of 196 samples (\ie, whole test set) by user turning on Chem dataset. 
Note that we set $\alpha=3.5$ because this is sufficient to achieve strong emotion conversions, as shown in Figure~\ref{fig:intensity_ratio}. 
The results are shown in Table~\ref{emo_speech_quality}. 
We find that the emotional controlling did not have much effect on the audio-visual alignment, and the lip sync of the seven emotions was still well expressed (see LSE-D and LSE-C). 
In addition, the WER of the six emotions (except ``sad'') is lower than 12.0\%, indicating the pronunciation clarity was not affected. 
As for ``sad'', it may be because the model does blur the pronunciation when imitating strong sadness. %
Finally, in terms of speaker similarity, the seven emotions are still very close to the original speech, which shows that emotion control does not affect the timbre of information.

\begin{table}[!tbp]
  \centering
  \resizebox{1.0\linewidth}{!}
  {
    \begin{tabular}{lcccccc}
    \toprule
    \# & Emotion   & LSE-C $\uparrow$ & LSE-D $\downarrow$ & WER $\downarrow$ & SECS $\uparrow$  \\
    \midrule
    0 & Neutral & 8.11 & 6.92 & \textbf{11.32} & 89.05 \\ 
    1 & Happy  & 8.10 &6.93  & 11.83 & 89.19 \\
    2 & Sad   & 8.06 &6.92  & 12.12 & 89.48 \\
    3 & Fearful & 8.09 & 6.90 & 11.91  & 89.00 \\
    4 & Surprise  & 8.09 & \textbf{6.89} & 11.96  & 87.91  \\
    5 & Disgusted  & 8.11 & 6.90 & 11.98  & 88.28  \\
    6 & Angry  & 8.10 & 6.94 & 11.76  & 88.63 \\
    \midrule
    7 & Original (w/o FUEC)  & \textbf{8.11}    &  {6.92}   &  {11.72} & \textbf{90.62}  \\ 
    \bottomrule
    \end{tabular}%
    } 
    \vspace{-6pt}
  \caption{Emotional speech quality study of EmoDubber. } 
  \vspace{-0.4cm}
  \label{emo_speech_quality}%
\end{table}

\noindent\textbf{Qualitative Results}. 
To demonstrate the effect of EmoDubber’s emotion synthesis, we invite three volunteers to express their expected emotions and intensities on three videos from dubbing datasets, respectively. 
We visualize the mel-spectrograms and provide comparison using FUEC or not. 
For instance, in (a), the user selects ``surprise'' (c=4) with an intensity of $\alpha=3.1, \beta=0.3$, which results in a noticeable rise trend at the end of the spectrograms (see blue box). 
In (b), the user selects ``angry'' (c=6) with an intensity of $\alpha=4.5, \beta=0.9$, which brings a remarkable increase in energy.  It indicates that a strong tone is being expressed to render angry. 
Finally, in (c), the user selects ``sad'' (c=2) with an intensity of $\alpha=5.0, \beta=1.7$. 
This relatively high-intensity setting produces a spectrum with diminished high-frequency energy and softened transitions, typical of frustrated expressions associated with sadness.

\vspace{-3pt}
\section{Conclusion}

In this work, we propose EmoDubber, a controllable emotion dubbing architecture to help users specify the emotion they need while satisfying high-quality lip sync and clear pronunciation. 
The LPA learns the inherent consistency by duration-level contrastive learning to reason the correct audio-visual alignment. 
Building on contextual information, the proposed PE strategy fuses the explicit phoneme chunk on video level to improve pronunciation by efficient conformer. 
Besides, the Flow-based User Emotion Controlling (FUEC) with positive and negative guidance dynamically adjusts the flow-matching prediction process to control emotion intensity flexibly.
Extensive experiments on real dubbing benchmarks show favorable performance. 

\section{Acknowledgement}

This work was supported by National Natural Science Foundation of China: 62322211, 61925201, 62132001, 62432001, and Beijing Natural Science Foundation (L247006), ``Pioneer'' and ``Leading Goose'' R\&D Program of Zhejiang Province (2024C01023), Key Laboratory of Intelligent Processing Technology for Digital Music (Zhejiang Conservatory of Music), Ministry of Culture and Tourism (2023DMKLB004). 
Yuankai Qi, Anton van den Hengel, and Jian Yang are not supported by the aforementioned fundings.

{
    \small
    \bibliographystyle{ieeenat_fullname}
    \bibliography{main}
}

\end{document}